\documentclass[prb, showpacs,preprint]{revtex4}
\usepackage{bm}
\usepackage{graphicx}%
\begin{document}
\title
{Locking and unlocking of the counterflow transport in $\nu=1$
quantum Hall bilayers by tilting of magnetic field}
\author{D.\,V.\,Fil$^{1,2}$ }
\affiliation{%
$^1$Institute for Single Crystals, National Academy of Sciences of
Ukraine, Lenin Avenue 60, Kharkov 61001, Ukraine\\ $^2$Max Planck
Institute for Solid State Research, Heisenbergstrasse 1, D-70569
Stuttgart, Germany}

\begin{abstract}
The counterflow transport in quantum Hall bilayers provided by
superfluid excitons is locked at small input currents due to a
complete leakage caused by the interlayer tunneling. We show that
the counterflow critical current $I_{c}^{\rm{CF}}$ above which the
system unlocks for the counterflow transport can be controlled by
a tilt of magnetic field in the plane perpendicular to the current
direction. The effect is asymmetric with respect to the tilting
angle. The unlocking is accompanied by switching of the systems
from the d.c. to the a.c. Josephson state. Similar switching takes
place for the tunneling set-up when the current flowing through
the system exceeds the critical value $I_c^{\rm{T}}$. At zero tilt
the relation between the tunnel and counterflow critical currents
is $I_c^{\rm{T}}=2 I_{c}^{\rm{CF}}$. We compare the influence of
the in-plane magnetic field component $B_\parallel$ on the
critical currents $I_{c}^{\rm{CF}}$ and $I_c^{\rm{T}}$. The
in-plane magnetic field reduces the tunnel critical current and
this reduction is symmetric with respect to the tilting angle. It
is shown that the difference between $I_{c}^{\rm{CF}}$ and
$I_c^{\rm{T}}$ is essential at field $|B_\parallel|\lesssim
\phi_0/d \lambda_J$, where $\phi_0$ is the flux quantum, $d$ is
the interlayer distance, and $\lambda_J$ is the Josephson length.
At larger $B_\parallel$ the critical currents $I_{c}^{\rm{CF}}$
and $I_c^{\rm{T}}$ almost coincide each other.
\end{abstract}


\maketitle

The idea on exciton superfluidity in electron-hole bilayers
\cite{1,2,3} and especially in  quantum Hall bilayers \cite{4,6,7}
with total filling factor  $\nu_T=1$ has obtained a lot of
attention in past ten years because of comprehensive experimental
study of that problem. In view of possible applications the most
important are the counterflow experiments\cite{8,9,10}. In these
experiments the samples with separate assess to the layers are
used.  Electrical current is injected into one layer in a given
end of the Hall bar, is withdrawn from the same layer in the
opposite end, and is redirected to the other layer. The currents
in the layers have the same value and opposite directions, so they
may be provided solely by superfluid magnetoexcitons.

Samples used in the counterflow experiments\cite{8,9,10}
demonstrate a huge increase of conductivity at low temperatures,
but they do not demonstrate zero counterflow resistance. We
consider that the zero-resistance state can be realized only in
quite perfect bilayers. Imperfectness results in emergence of
vortices (merons) in the magnetoexciton gas. Meron local
concentration is proportional to the deviation of the local
filling factor from unity. At rather strong imperfectness merons
become uncoupled at all temperatures and their motion
perpendicular to  the charge transport direction results in a
finite counterflow resistance\cite{m1, m2, m4}. At low degree of
imperfectness meron pairs remain bounded and the counterflow
resistance should go to zero.

Magnetoexciton superfluidity in bilayers is possible at rather
small interlayer separation $d$ (less or of order of the magnetic
length $\ell_B$). At such a separation the interlayer tunnelling
is not negligible and it may influence significantly on the
counterflow transport\cite{16,17,18,30,t4}. This influence is
connected with a formation of another type of vortices - the
Josephson ones. The length parameter associated with Josephson
vortices is the Josephson length
$\lambda_J=\ell_B\sqrt{2\pi\rho_s/t}$, where $t$ is the interlayer
tunneling amplitude, and $\rho_s$ is the superfluid stiffness for
magnetoexcitons. If $\lambda_J$ is much smaller than the length of
the Hall bar $L_x$, the effect of locking of the bilayer for the
counterflow transport takes place.

The locking occurs at small input current $I_{in}<I_c^{\rm{CF}}$
at which a partial Josephson vortex is formed at the source end
and
the current does not reach the load end. The input critical
current is equal to the integral Josephson current for the half of
the Josephson vortex: $I_c^{\rm{CF}}= 2 j_{0} \lambda_J L_y$,
where $j_0=e t/2\pi\hbar\ell_B^2$ is the maximum Josephson current
density, and $L_y$ is the width of the Hall bar. One can see that
in this state, that is a kind of the d.c. Josephson
state\cite{dc}, the integral Josephson current is proportional to
$\sqrt{t}$.
 At $I_{in}>I_c^{\rm{CF}}$  the Josephson vortex chain emerges
instead of the  partial vortex, and the current reaches the load.
Nonzero current in the load circuit requires nonzero interlayer
voltage. This voltage forces the vortex chain to move along the
bilayer. Such a state is a kind of the a.c. Josephson state. In
this state the leakage is small - the integral average in time
Josephson current is proportional to $t^2$. A finite value of this
current is connected with dissipative processes  that switch on
\cite{16,18} in the a.c. state.
 The effect of locking and
unlocking of the quantum Hall bilayer for the counterflow
transport was observed in the recent experiment\cite{19}.

In view of possible applications of the exciton superfluidity it
is important to control the locking-unlocking effect. In this
paper we show that the in-plane component of magnetic field
$B_\parallel$ can be used for such a control. We have found that
the dependence of $I_c^{\rm{CF}}(B_\parallel)$  is asymmetric and
one can decrease or increase the critical current by tilting.

We restrict our study with the case of perfect bilayers without
free merons and do not consider the influence of the meron induced
disorder\cite{t1,t5} on the critical current.

The switching  between the d.c and a.c. Josephson regimes takes
place in another experimental set-up called the tunneling one and
used for the observation of the Josephson effect in bilayers
\cite{11,12,13,14,15}. In this set-up the current is injected into
the top layer at one end of the Hall bar and is withdrawn from the
bottom layer at the opposite end.
 In the d.c. Josephson state two
partial Josephson vortices are formed at the both ends of the Hall
bar and normal co-directed intralayer currents flow in the bulk.
At zero in-plane magnetic field the maximum current in the d.c.
state is $I_c^{\rm{T}}=2 I_c^{\rm{CF}}$ (the factor of 2 is due to
additive contribution of two ends of the Hall bar). The interlayer
voltage in the d.c. state is equal to zero. The transition from
the d.c. to the a.c. Josephson regime  reveals itself in a sharp
drop of the intergal Josephson current. The value of the tunnel
critical current can be extracted from the I-V characteristics:
the maximum current before its drop is identified as
$I_c^{\rm{T}}$ \cite{14,15}. We would note that in the
experiments\cite{14,15} the voltage between two leads in the
central part of the sample was much smaller than the voltage
between  the input and the output leads. It probably means that
the bias voltage measured in tunneling experiments is mostly the
contact voltage.

In the tunneling set-up the in-plane magnetic field may cause a
resonant increase of the integral tunnel current in the a.c.
regime\cite{t1,t2,t3} (similar behavior was also observed
experimentally\cite{12}). In view of the exact relation between
$I_c^{\rm{T}}$ and $I_c^{\rm{CF}}$ at zero tilt it is of interest
to consider how the tilt changes the critical current
$I_c^{\rm{T}}$. We find that in balanced bilayers the function
$I_c^{\rm{T}}(B_\parallel)$ is symmetric and the tilt
(irrespective to its sign) results in a decrease of the tunnel
critical current.

We will formulate the problem in terms of the phase of the order
parameter $\varphi$ for the superfluid magnetoexciton gas. The
axis $x$ is chosen along the flow direction and the derivative $d
\varphi/d x$ determines the intralayer supercurrents
\begin{equation}\label{2}
    j_{s1}=-j_{s2}=\frac{e}{\hbar}\rho_s\left( \frac{d
  \varphi}{d x}-\frac{e B_y d}{\hbar c}\right).
\end{equation}
Here and below we imply that $|d \varphi/d x|\ll \ell^{-1}_B$. We
specify the case  of the phase $\varphi$ independent of $y$ and
the magnetic field tilted in the plane perpendicular to the
current direction ($B_\parallel=B_y$). The Josephson current
density reads as
\begin{equation}\label{3}
j_{J}=-\frac{e}{\hbar}\frac{t}{2\pi \ell^2_B}\sin\varphi.
\end{equation}
The quantity $j_J$ is defined as a current that flows from the
layer 1 to the layer 2.  The intralayer currents contain the
uniform counterflow diamagnetic component
\begin{equation}\label{dd}
    j_d=-\frac{e^2 \rho_s B_y d}{\hbar^2 c}.
\end{equation}
The diamagnetic effect is rather small:  the magnetic
susceptibility $\chi=-(e^2/\hbar c)^2 \rho_s d/e^2$ is
proportional to the square of the fine structure constant.
Therefore the difference between the external magnetic field and
the field inside the bilayer can be neglected. But the presence of
the diamagnetic current is significant for the transport
properties.

In the d.c. state the local interlayer voltage is equal to zero
($V_1({\bf r})=V_2({\bf r})$) that means the equivalence of
electrical fields in the layers (${\bf E_1}={\bf E_2}={\bf E}$).
The currents satisfy the stationary continuity equations $d
j_{1(2)}/d x \pm j_{J}=0$, where the intralayer current is the sum
of the supercurrent and the normal current
($j_{1(2)}=j_{s1(2)}+j_{n1(2)}$). Taking into account the
condition $j_{s1}=-j_{s2}$, one finds that ${\bf j}_{n1}+{\bf
j}_{n2}=(\hat{\sigma}_1+\hat{\sigma}_2){\bf E}=const$, where
$\hat{\sigma}_i$ is the normal conductivity tensor for the layer
$i$.

In the d.c. state the current in the load circuit should be zero
(in the counterflow set-up). Therefore $j_1(L_x)=j_2(L_x)=0$ that
yields $j_{n1}(L_x)+j_{n2}(L_x)=0$. Thus $j_{n1}+j_{n2}=const=0$,
the electrical field ${\bf E}=0$, and $j_{n1}=j_{n2}=0$. The
continuity equations are reduced to the following equation for the
phase
\begin{equation}\label{4}
 \frac{d^2 \varphi}{d x^2}= \frac{1}{\lambda_J^2}\sin \varphi.
\end{equation}
Eq. (\ref{4}) is the nonlinear pendulum equation in which the time
variable is replaced with the space one. Two different types of
motion of a nonlinear pendulum (oscillation and rotation)
correspond to two distinct d.c. Josephson states. They are
classified as the vortex-antivortex (VA) chain, and the vortex (V)
(or the antivortex (A) chain) state.  The word "vortex"
("antivortex") stands  for the Josephson vortices  with the
positive (negative) vorticity.

The currents in the VA state have the form
\begin{eqnarray}\label{5}
  j_{s1}(x)=j_d + j_c \sqrt{\eta} \, {\rm cn}\left(\frac{x-x_0}{\lambda_J},
  \eta\right),\cr
   j_{J}(x)= \frac{j_c  \sqrt{\eta}}{\lambda_J} \,
  {\rm dn}\left(\frac{x-x_0}{\lambda_J},
  \eta\right) {\rm sn}\left(\frac{x-x_0}{\lambda_J},
  \eta\right).
\end{eqnarray}
The V (A) state configuration of currents is described by the
equation
\begin{eqnarray}\label{6}
   j_{s1}(x)=j_d\pm \frac{j_c}{\sqrt{\eta}}
  {\rm dn}\left(\frac{x-x_0}{\lambda_J\sqrt{\eta}},
  \eta\right),\cr
   j_{J}(x)= \pm\frac{j_c  }{\lambda_J}  {\rm sn}
   \left(\frac{x-x_0}{\lambda_J\sqrt{\eta}},
  \eta\right) {\rm cn}\left(\frac{x-x_0}{\lambda_J\sqrt{\eta}},
 \eta\right).
 \end{eqnarray}
In Eqs. (\ref{5}), (\ref{6}) $j_c=2
{e}{\rho_s}/{\lambda_J}{\hbar}$, is the critical current density,
and ${\rm sn}(x,\eta)$, ${\rm cn}(x,\eta)$ and ${\rm dn}(x,\eta)$
are the Jacobi elliptic functions. The parameter $\eta$ is in the
range $(0,1]$. This parameter is connected with the period of the
vortex chain. At $\eta\to 1$ the period  goes to infinity, and Eq.
(\ref{5}), as well as Eq. (\ref{6}), describes a single vortex
centered at $x_0$.

The energy of the Josephson vortex state is given by the equation
\begin{equation}\label{1}
  E=\int d^2 r \left[ \frac{1}{2} \rho_s  \left( \frac{d
  \varphi}{d x}-\frac{e B_y d}{\hbar c}\right)^2-
  \frac{{t}}{2\pi \ell^2_B} \cos
  \varphi\right].
\end{equation}
The conditional minimum of the energy (\ref{1}) at given boundary
conditions for the input and output currents determines the vortex
configuration.

Prior to consider the critical current problem we would remind
that Josephson vortices can emerge at zero input current, as
well\cite{20,21}. If
$$|B_y|>B_c=\frac{4\phi_0}{\pi^2 d\lambda_J}$$
($\phi_0=hc/2e$ is the flux quantum) Josephson vortices penetrate
into the bulk of an isolated bilayer
($j_{1(2)}(0)=j_{1(2)}(L_x)=0$) and a vortex chain structure with
the period of order of $\lambda_J$ is formed. The in-plane
critical field $B_c$ is analogous to the the critical field
$H_{c1}$ for a long Josephson contact between two superconductors.
At $|B_y|\leq B_c$ a state with only two partial vortices situated
at the opposite ends of the Hall bar is realized. These partial
vortices joint counterflow diamagnetic intralayer currents into
the circular diamagnetic current.

For the counterflow set-up the critical current density
$j_c^{\rm{CF}}(B_y)$ can be found as  follows. In the d.c. state
the normal current is equal to zero.  Thus $j_{s1}(0)=j_{in}$,
where $j_{in}$ is the input current density, and $j_{s1}(L_x)=0$.
We imply that the phase $\varphi(x)$ is a continuous function of
$x$. It corresponds to the vortex state with the same $\eta$ in
the whole system. For a given state the intralayer current varies
in a certain range determined by the parameter $\eta$ and the
in-plane field $B_y$. It is the range $[j_d-j_c
\sqrt{\eta},j_d+j_c \sqrt{\eta}]$ for the VA state, the range
$[j_d+j_c \sqrt{(1-\eta)/\eta},j_d+j_c \sqrt{1/\eta}]$ for the V
state, and the range $[j_d-j_c \sqrt{1/\eta},j_d-j_c
\sqrt{(1-\eta)/\eta}]$ for the A state. The counterflow d.c. state
can be realized if at least one of the ranges enumerated above
contains both $j_{in}$ and zero. The situation is symmetric with
respect to the change of sign of both $j_{in}$ and $B_y$, and it
is enough to consider only positive $j_{in}$.

For the further analysis it is convenient to define the
characteristic in-plane magnetic field
\begin{equation}\label{cf}
   B'_{c}=\frac{2}{\pi}\frac{\phi_0} {d\lambda_J}
\end{equation}
determined by the condition $|j_d|=j_c$. It is larger than $B_c$
($B'_{c}=\pi B_c/2$).

Let us first consider the case of positive $B_y$ ($j_d<0$). The VA
state may satisfy the boundary conditions if $j_{in}\leq
j_c-|j_d|$.  The V state is possible if for the same $\eta$ two
inequalities $j_{in}<j_c \sqrt{1/\eta}-|j_d|$ and $j_c
\sqrt{(1-\eta)/\eta}-|j_d|<0$ are fulfilled. That yields the
following restriction on the input current: $j_{in}\leq
\sqrt{j_c^2+j_d^2}-|j_d|$.

At  $-B'_c<B_y<0$ ($j_d>0$) the d.c. state is possible up to
$j_{in}=j_d+j_c$. Indeed, the VA state satisfies the boundary
conditions at $j_d<j_{in}\leq j_d+j_c$, and the A state - at
$j_{in}\leq j_d$. At $B_y>-B'_c$ the VA state is not possible, and
the A state may satisfy the d.c. boundary conditions only for the
input current $j_{in}<j_d-\sqrt{j_d^2-j_c^2}$.

\begin{figure}
\begin{center}
\includegraphics[width=8cm]{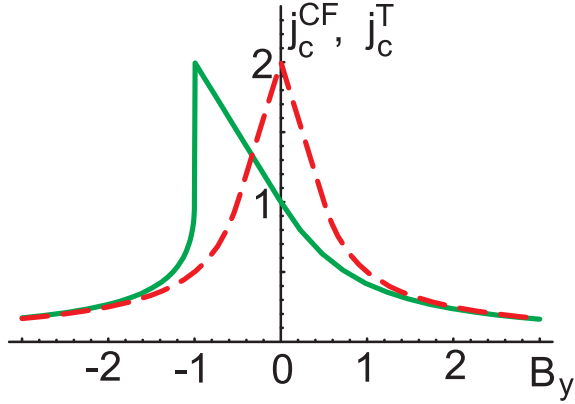}
\end{center}
\caption{Critical current densities (in $j_c$ units) vs. the
in-plane magnetic field (in $B'_c$ units). The counterflow
critical current $j_c^{\rm{CF}}$ is shown by solid line, the
tunnel critical current $j_c^{\rm{T}}$ - by dashed line.}
\label{f1}
\end{figure}

Thus the dependence of the counterflow critical current density on
$B_y$ has the form
\begin{equation}\label{11}
    j_c^{\rm{CF}}(B_y)=j_c\cases{-\frac{B_y}{B'_c}
    -\sqrt{\left(\frac{B_y}{B'_{c}}\right)^2-1}
  & at $B_y<-B'_{c}$
  \cr
  1-\frac{B_y}{B'_{c}}& at $-B'_{c}\leq B_y<0$
  \cr
  \left[\sqrt{1+\left(\frac{B_y}{B'_{c}}\right)^2}
  -\frac{B_y}{B'_{c}}\right]& at
  $B_y>0$}
\end{equation}
The dependence (\ref{11}) is shown in Fig. \ref{f1}. One can see
that at $|B_y|\lesssim B'_c$ this dependence is essentially
asymmetric one. Such an asymmetry is connected with that the
counterflow current and the diamagnetic current can be co-directed
or oppositely directed depending on the sign of the tilting angle.
The  tilting angle that corresponds to $B_y=B'_c$ is rather small:
$\phi^c_{\rm tilt}\approx{2\ell^2_B}/({d \lambda_J})$.

Let us say some words on the role of critical field $B_c$. The
dependence $j_c^{\rm{CF}}(B_y)$ is continuous one at $|B_y|=B_c$.
But the vortex structure that corresponds to the energy minimum
changes significantly at this point. At $|B_y|<B_c$ the V and A
states are the states with only two partial vortices
(antivortices) at the opposite ends. But if $|B_y|$ exceeds $B_c$
these states are transformed into multivortex ones. The VA state
with minimal energy is the state with only a partial vortex at one
end and a partial antivortex at the opposite end irrespective of
the value of $B_y$.

Let us now switch to the tunneling set-up. In this set-up
$j_1(0)=j_2(L_x)=j_{in}$ and $j_2(0)=j_1(L_x)=0$. Since the
counterflow currents cannot transfer the charge between two ends,
 normal currents are nonzero and their sum is equal to the input current
${j}_{n1}+{ j}_{n2}=j_{in}=const$. The difference ${j}_{n1}-{
j}_{n2}=const$,  as well. Thus the normal current does not enter
into the continuity equation, and the latter is reduced to the
equation for the phase (\ref{4}).

Here we specify the case of balanced bilayers in which ${
j}_{n1}={ j}_{n2}$, and the supercurrents satisfy the boundary
conditions $j_{s1}(0)=-j_{s1}(L_x)=j_{in}/2$. For given $B_y$ and
$\eta$ we have  three ranges of $j_{s1}$ (that coincide with ones
given above) for the VA, V and A solutions.  In the tunneling
set-up the d.c. state can be realized if the quantities
$+j_{in}/2$ and $-j_{in}/2$ belong to the same range.  For the VA
solution the latter condition is fulfilled under the following
restriction on the value of the input current
\begin{equation}\label{10-1}
    |j_{in}|<2(j_c-|j_d|).
\end{equation}
The V solution may satisfy the boundary condition at negative
$j_d$, and the A solution - at positive $j_d$. Common for both
solutions restriction on $j_{in}$ reads as
\begin{equation}\label{11-1}
    |j_{in}|<\max[ F(\eta)],
\end{equation}
where the function
\begin{equation}\label{11-1-1}
 F(\eta)= 2\min(|j_d|-j_c\sqrt{\frac{1-\eta}{\eta}},
    j_c\frac{1}{\sqrt{\eta}}-|j_d|)
\end{equation}
contains $B_y$ as a parameter and is defined in the interval
$0<\eta\leq 1$. Let us find $\eta_m$ that maximizes the function
$F(\eta)$. At $|j_d|<j_c/2$ we obtain $\eta_m=1$ and
$F(\eta_m)=2|j_d|$. At $|j_d|>j_c/2$ the quantity $\eta_m$ is
determined by the equation
\begin{equation}\label{12-1}
    |j_d|-j_c\sqrt{\frac{1-\eta_m}{\eta_m}}=
    j_c\frac{1}{\sqrt{\eta_m}}-|j_d|
\end{equation}
that yields $\sqrt{\eta_m}=4 j_c |j_d|/(4j_d^2+j_c^2)$ and
$F(\eta_m)=j_c^2/2|j_d|$

Comparing the conditions (\ref{10-1}) and (\ref{11-1}) we obtain
the final expression for the tunnel critical current density
 \begin{equation}\label{14-1}
    j_c^{T}(B_y)=2j_c \cases{1-\frac{|B_y|}{B'_c}
  & at $|B_y|<B'_{c}/2$
  \cr
  \frac{B'_c}{4 |B_{y}|}& at $|B_y|\geq B'_{c}/2$}
\end{equation}
The dependence (\ref{14-1}) is shown in Fig. \ref{f1}. One can see
that while at $B_y=0$ the current $j_c^{T}$ exceeds $j_c^{CF}$ by
the factor of two, at $|B_y|\gg B'_c$ these quantities almost
coincide each other. The other difference between $j_c^{T}$ and
$j_c^{CF}$ is that the tunneling critical current is symmetric
with respect the sign of the tilting angle. The latter property
can also be predicted from the symmetry reasons. Note that such a
symmetry takes place only in case of balanced bilayers: at nonzero
imbalance ${ j}_{n1}\ne { j}_{n2}$, that results in asymmetric
dependence $j_c^{T}(B_y)$.

In conclusion, we have shown that the locking and unlocking of the
quantum Hall bilayer for the counterflow transport can be
controlled by tilting of magnetic field. The effect can be
observed in the same experimental set-up, where the
locking-unlocking effect under variation of the input current was
recently discovered\cite{19}. Asymmetric dependence of the
critical current on magnetic field is expected in a rather narrow
diapason of tilting angles close to zero. We have compared the
influence of the in-plane magnetic field on the counterflow
critical current and on the tunnel critical current\cite{14,15}.
We find that the difference is essential at small in-plane
magnetic fields. The maximum counterflow critical current
coincides with the maximum tunnel critical current, but in the
first case the maximum is reached at $B_y=-B'_c$, while in the
second case - at $B_y=0$.

It is important to discuss the validity of our results for real
experimental systems. The  main assumption of our consideration is
the existence of a path between the input and the output end that
is free from merons and weak links. We imply that the phase of the
order parameter is continuous one along this path. Systems, where
such a path does not exist, but which have quite long areas
without merons may also demonstrate similar behavior. In the
latter case the tunnel critical current at $B_y\ne 0$ should be
larger than in the case considered in this paper. It is because
two ends will work separately.

This study was supported by the Stipendium der Max Planck
Gesellschaft.


\begin{thebibliography}{99}
\bibitem{dc}
Here the d.c. state is understood as a state in which
 the current does not contain hf Josephson-type oscillations.
 It takes place for a d.c. input current, as well as, for a low
 frequency a.c. input
 current,
 if its amplitude is lower than the critical
 one.
\end{thebibliography}
\end{document}